\documentstyle[prl,aps,preprint]{revtex}
\def\Av{{\bf A}}

\def\ev{{\bf e}}

\def\Hv{{\bf H}}

\def\Nv{{\bf N}}

\def\Rv{{\bf R}}

\def\tv{{\bf t}}

\def\xv{{\bf x}}

\def\Aa{{\cal A}}
\def\Ha{{\cal H}}

\def\gradv{{\mbox{\boldmath{$\nabla$}}}}

\def\sigmav{{\mbox{\boldmath {$\sigma$}}}}

\def\rhob{{\overline \rho}}
\def\ssv{\sigmav}
\def\ss{\sigma}
\def\gb{{\overline g}}
\def\Kb{{\overline K}}
\def\Oom{\Omega}
\def\Ac{A}

\begin{document}

\widetext
\draft

\title{
$n$-atic Order and Continuous Shape Changes of Deformable Surfaces of Genus
Zero
}
\author{Jeong-Man Park and T. C. Lubensky}
\address{
Department of Physics, University of Pennsylvania, Philadelphia, PA 19104}
\author{F. C. MacKintosh}
\address{
Department of Physics, University of Michigan, Ann Arbor, MI 48109}
\maketitle

\begin{abstract}
We consider in mean-field theory the continuous development below a 
second-order phase transition of $n$-atic tangent plane order 
on a deformable surface of genus zero with
order parameter $\psi = \langle e^{i n \theta} \rangle$.  Tangent plane
order expels Gaussian curvature. 
In addition, the total vorticity of orientational order on a
surface of genus zero is two.  Thus, the ordered phase of an $n$-atic
on such a surface
will have $2n$ vortices of strength $1/n$, $2n$ zeros in its order
parameter, and a nonspherical equilibrium shape.  
Our calculations are based on a
phenomenological model with a gauge-like coupling between $\psi$ and
curvature, and our analysis follows closely the Abrikosov treatment of a
type II superconductor just below $H_{c2}$. 
\end{abstract}
\par
When aliphatic molecules are dissolved in water, they  can
spontaneously
segregate into bilayer membranes in which hydrocarbon tails are shielded from
contact with water.  Depending on conditions, these membranes can form random,
extended surfaces, regular periodic structures, or closed vesicles
separating an interior from an exterior[1].  Membranes of a similar nature but
with more structure and
greater complexity form biological cell walls.
In a wide class of membranes
molecules move freely within the membrane forming a $2$-dimensional
fluid offering little resistance to changes in membrane shape. 
Such membranes are physical examples of random surfaces, which
can undergo violent shape changes.
Molecules in membranes can also exhibit
varying degrees of orientational and positional order[2] including tilt,
hexatic, and crystalline order similar to that found in free standing liquid 
crystal
films [3].  These ordered membranes provide fascinating laboratories for
the study of the coupling between order and geometry, analogous in many ways
to the coupling between matter and geometry in general relativity.  The
underlying cause of this coupling is easy to see.  A vector field (or any
field describing orientational order) cannot be everywhere parallel if it is
forced to lie on a surface, such as
that of a sphere, that curves in two directions and thus has nonzero
Gaussian curvature.  It could lower its energy
by flattening the surface.  An additional complication arises when order
develops on closed surfaces.  A closed surface can be classified according
to its genus (or number of handles)
$g$: a sphere has genus zero, a torus genus one, etc.
Orientational order on a closed surface necessarily [4] has topological
defects (vortices) with total strength (vorticity) equal to the Euler
characteristic $2(1-g)$ of the surface.  Tangent plane order on a
sphere will have vorticity $2$, a torus vorticity $0$, etc.  
The continuous development of vector order on a deformable surface of genus
zero will be accompanied by a continuous change from spherical to
ellipsoidal shape[5].
Since vortices
are energetically costly, it may be favorable for a closed physical
membrane to transform into an open cylindrical structure [6] when tangent
plane order develops in response to changes in temperature or other control
variable.  Indeed, there are a number of experimental examples of shapes
changes[7] that may
be explained by the development of tangent plane order.
\par
In this paper, we investigate in mean-field theory the development of
$n$-atic order on a closed surface of genus zero and the concomitant
change in shape from spherical to non-spherical.  An $n$-atic order
parameter can have vortices of strength $1/n$, and, since it is generally
favorable to form vortices of minimum strength, we expect $2n$ maximally
separated vortices of strength $1/n$ to be present in the ordered phase.  Thus
for $n=1$, we expect two antipodal vortices[5], and respectively
for $n=2$, $n=3$ and $n=6$, we expect 
vortices at the vertices of a tetrahedron, an octahedron, and an icosahedron.
Indeed calculations on a
rigid sphere confirm this conjecture[6,8].
We also expect the vesicle
shape to change from spherical to ellipsoidal, tetrahedral, octahedral, or
icosahedral in the four cases above.  Fig. 1 shows our calculated shapes 
for temperatures just below the transition temperature to the $n$-atic state
for $n=1,2,3,4$, and $6$.
\par
Our calculations are based on a phenomenological Hamiltonian[9] for a
complex order parameter field whose coupling to shape occurs via a
covariant derivative and via changes in the metric tensor.  The model is
almost identical to the Landau-Ginzburg theory of superconductivity except
that vorticity is fixed by surface topology rather than energetically
determined by an external magnetic field.  The ordering transition we find
is very similar to the transition from a normal metal to the Abrikosov
vortex lattice[10] in a superconductor, and indeed our analysis follows
very closely that of Abrikosov.  
We find the order parameter, which is
necessarily spatially inhomogeneous because of the topological constraint on
the total vorticity, from among a highly degenerate
set of functions that diagonalize the harmonic Hamiltonian on a rigid
sphere.  This degenerate set has exactly $2n$ zeros at arbitrary positions
on the sphere and is very similar in form to fractional quantum Hall wave
functions[11].
\par
Positions on a two-dimensional surface in $R^3$ are specified by a vector
$\Rv ( \ssv )$  as a function of a two-dimensional coordinate $\ssv = (\ss_1
, \ss_2 )$.  Associated with $\Rv ( \ssv )$ is a metric tensor 
$g_{ab} ( \ssv ) = \partial_a \Rv ( \ssv ) \cdot \partial_b \Rv ( \ssv )$
and a curvature tensor $K_{ab} ( \ssv )$ defined via $\partial_a \partial_b
\Rv = K_{ab} \Nv$ where $\Nv$ is the local unit normal to the surface.
To describe tangent-plane $n$-atic order, we  introduce
orthonormal unit vectors $\ev_1$ and $\ev_2$ at each point on the surface.
If $\tv ( \ssv )$ is any unit tangent vector, then $\ev_1 ( \ssv ) \cdot \tv
( \ssv ) = \cos \theta ( \ssv )$ defines a local angle $\theta ( \ssv )$.
$n$-atic order is then described by the order parameter $\psi ( \ssv ) = 
\langle e^{i n \theta ( \ssv )} \rangle$, 
which can be related to the $n$th rank symmetric
traceless tensor constructed from the vector $\tv$.  Note that since $\theta
( \ssv )$ depends on the choice of orthonormal vectors $\ev_1$ and $\ev_2$,
the order parameter $\psi( \ssv )$ does as well.  This means that any
spatial derivatives in a phenomenological Hamiltonian for $\psi$ must be
covariant derivatives.  The long-wavelength Landau-Ginzburg Hamiltonian for
$\psi$ is $\Ha_{\psi} = \Ha_1 + \Ha_2$ where
\begin{equation}
\Ha_1 = \int d^2 \ss  \sqrt{g} [ r | \psi |^2 + {1 \over 2} u | \psi
|^4 ] , \qquad 
\Ha_2  =  C \int d^2 \ss \sqrt{g} g^{ab} ( \partial_a - i n \Ac_a
) \psi ( \partial_b + i n \Ac_b ) \psi^* ,
\end{equation}
where $\Ac_a = \ev_1 \cdot \partial_a \ev_2$ is the spin connection, 
$g = \det g_{ab}$ and $g^{ab}g_{bc} = \delta_c^a$.  
When $n=6$, this $\Ha_{\psi}$ is identical to the
Hamiltonian for hexatic order on a deformable surface introduced in Ref.
[12].
It is important to note that the covariant derivative $D_a = \partial_a - i
n \Ac_a$ depends on $n$.
The energy associated
with shape changes of a constant area membrane is described by the Helfrich
Hamiltonian[13], 
\begin{equation}
\Ha_{\rm curv} = {1 \over 2} \kappa \int d^2 \ss \sqrt{g} ( K_a^a )^2 ,
\end{equation}
The complete
Hamiltonian for $n$-atic order on a deformable surface is 
$\Ha = \Ha_{\psi}+ \Ha_{\rm curv}$.
\par
We can now specialize to surfaces of genus zero with constant area 
$\Aa = 4 \pi R^2$ whose 
shapes do not differ significantly from that of a sphere. We choose $\ssv =
( \theta , \phi ) \equiv \Oom$ to be the polar coordinates of a sphere and set 
$\Rv ( \Oom ) = R_0 [ 1 + \rho ( \Oom )] \ev_r$,
where $\ev_r$ is the radial unit vector.  This allows us to introduce
reduced metric and curvature tensors $\gb_{ab}$ and $\Kb_{ab}$ via
$g_{ab} = R_0^2 \gb_{ab}$ and $K_{ab} = R_0^{-1} \Kb_{ab}$. The field $\rho (
\Oom )$ measures
deviations from sphericity and can be expanded in spherical
harmonics.  Any isotropic change in $\Rv$ can be described by $R_0$. 
In addition, uniform
translations, which change neither the shape or the energy of the vesicle,
correspond to distortions in $\rho$ with $l=1$.  These considerations imply
that $\rho$ will have no $l=0$ or $l=1$ components:
$\rho ( \Oom ) = \sum_{l=2}^{\infty}\sum_{m= -l}^l \rho_{lm} Y_l^m ( \Oom )$.
The shape and size of the vesicle are determined entirely by the parameters
$R_0$ and $\rho_{lm}$.  
The reduced tensors $\gb_{ab}$ and $\Kb_{ab}$ do not
depend on $R_0$. Therefore, $R_0$ can be expressed as a function of $\Aa$
and $\rho_{lm}$ via the 
relation $\Aa = \int d \Oom \sqrt{g} = R_0^2 \int d \Oom \sqrt{\gb}$. 
In the disordered phase $\rho = 0$, and $R = R_0$. 
We will use the Hamiltonian $\Ha = \Ha_{\psi} + \Ha_{\rm curv}$
expressed in terms of reduced parameters and the constant
area $\Aa$ in our calculations 
of shape changes below the second-order disordered-to-$n$-atic
transition.
\par
The similarities between $\Ha$ and the Landau-Ginzburg Hamiltonian for a
superconductor in an external magnetic field,
\begin{equation}
\Ha_{\rm LG} = \int d^3 x [ r | \psi |^2 + C | ( \gradv - 
ie^* \Av )\psi|^2
+ {1 \over 2 } u | \psi |^4 + {1 \over 8 \pi} ( \gradv \times \Av - \Hv )^2
],
\end{equation}
are striking
(Here $e^* = 2 e / \hbar c$).  Both have a complex order
parameter $\psi$ with covariant derivatives providing a coupling between
$\psi$ and a ``vector potential" $\Av$ or $\Ac_a$.  In a magnetic field, the
superconductor can undergo a $2$nd order mean-field transition from a normal
metal to the Abrikosov vortex lattice phase with a finite density of
vortices determined energetically by temperature and the magnetic field
$\Hv$.  The magnetic field is conjugate to the vortex number $N_{\rm v}$
since $\int d^3 x ( \gradv \times \Av ) = L N_{\rm v} \phi_0$, where $L$ is
the length of the sample along $\Hv$ and $\phi_0 = hc / 2e$ is the flux quantum.
On a closed surface with $n$-atic order, 
there is a second-order mean-field transition to a
state with vortex number determined by topology rather than conjugate
external field.  Thus, the $n$-atic transition on a closed surface is
analogous to transition to an Abrikosov phase with a fixed number of
vortices rather than fixed field conjugate to vortex number.  
\par
Before proceeding with our analysis of the $n$-atic transition on a
deformable sphere, it is useful to recall Abrikosov's calculation of the
transition to the vortex state.  The first step is to calculate the
eigenfunctions of the Harmonic part of $\Ha_{\rm GL}$ when $\gradv \times
\Av = \Hv$.  These can be divided into highly degenerate sets separated by
an energy gap $\hbar \omega_c = 2C e^* H$.  In the Landau gauge with $\Av = 
(0,Hx,0)$, the eigenfunctions in the lowest energy manifold are $\psi_k =
e^{iky} e^{- e^* H( x - x_k )^2 }$ where $x_k = k/ e^* H$.  The order
parameter $\psi ( \xv )$ of the ordered state is expressed as a linear
combination $\psi ( \xv ) = \sum C_k \psi_k$ where the complex parameters
$C_k$ are determined by minimization of $\Ha_{\rm GL}$
\par
We will now proceed to analyze our problem in the same spirit.  We first
diagonalize $\Ha_2$ in the high-temperature spherical state when
$\rho = 0$ and $\Ac_{\phi} = \Ac_{\phi}^0= - \cos \theta$, 
$\Ac_{	\theta } = \Ac_{\theta}^0 = 0 $, i.e., we
determine the functions $\psi$ that satisfy
$C D_a D^a \psi = \epsilon \psi$,
for $\rho = 0$ $\Ac_a = \Ac_a^0$.
In the lowest energy manifold, $\psi$ will have exactly $2n$ zeros
specifying the positions of vortices of strength $1/n$.  
In the stereographic projection
gauge where $z = \tan ( \theta / 2 ) e^{i \phi}$ , we find
\begin{equation}
\psi = \alpha \left( \frac{2|z|}{1+|z|^{2}} \right)^{n} \frac{1}{z^{n}}
       \prod_{i=1}^{2n}(z-z_{i}),
\label{wavefn1}
\end{equation}
and $\epsilon = C n / R$.
This function has zeros at the $2n$ points $z_i$ and only at these points.
Since these functions are polynomials in $z$ multiplied by a common
prefactor, any linear combination of them will yield another function of
exactly the same form but with different positions of zeros.  Thus, Eq.
(\ref {wavefn1}) is the most general function in the lowest energy manifold.
We note the similarity between $\psi^{1/n}$ and fractional quantum Hall
wavefunctions[11].  Functions in higher energy manifolds will have more zeros
corresponding to the creation of $\pm$ vortex pairs.
Reexpressing $z$ and $z_i$ in polar coordinates and choosing $\alpha$
appropriately, we obtain
\begin{equation}
\psi = \psi_0
       \prod_{i=1}^{2n} \left(
       \sin\frac{\theta}{2}\cos\frac{\theta_{i}}{2}e^{i(\phi-\phi_{i})/2} 
- \cos\frac{\theta}{2}\sin\frac{\theta_{i}}{2}e^{-i(\phi-\phi_{i})/2}
       \right) \equiv \psi_0 P ( \Oom )
\label{wavefn2}
\end{equation}
as a function of the polar coordinates $\Oom_i = ( \theta_i , \phi_i )$ of
the positions of its zeros on a sphere.
\par
Since $\psi$ of Eq. (\ref{wavefn2}) 
is the most general function in the lowest energy
manifold, the order parameter and vesicle shape (determined by $\rho$) just
below the transition are obtained by minimizing $\Ha$ over $\psi_0$
and the positions of zeros.  To order $\psi^4$, we can write $\Ha$ as
\begin{equation}
\Ha  =  {\Aa \over 4 \pi}
\int d \Omega  \left[( r - r_c ) 
| \psi |^2 + {1 \over 2 } 
u | \psi |^4 \right]  
+ \int d \Omega \rho ( \Omega ) \Phi '( \Omega ) + {\kappa \over 2} \sum l (
l^2 -1 )( l +2 ) | \rho_{lm} |^2 , \nonumber
\end{equation}
where  $r_c = -4 \pi n C/\Aa$ and
\begin{equation}
\Phi ' ( \Omega )  =
( \Aa / 2 \pi ) ( r | \psi |^2 + {1 \over 2 } u | \psi |^4 )
- {1 \over \sin \theta } \partial_a ( \gamma_b^a J^b ) 
 \equiv  \psi_0^2 \Phi ( \Oom )
\end{equation}
with $J^b = g^{-1/2}\delta \Ha / \delta \Ac_b$ evaluated at 
$\rho_{lm} =0$ and $R_0 = R$.  
Minimization over $\rho$ and $\psi_0$ leads
to $\psi_0^2 = - r[\{\Oom_i\}]/u[\{\Oom_i\}]$, 
\begin{equation}
\rho_{lm}  =  - {1 \over \kappa l ( l^2 -1 ) (l+2)} \int d \Omega Y_{lm}^* 
( \Oom )\Phi ' ( \Oom ) 
\equiv  - {\psi_0^2 \over \kappa l ( \l^2 - 1 ) (l+2)} \Phi_{lm} ( \Oom ) ,
\end{equation}
and the effective free energy density, 
\begin{equation}
f [\{\Oom_i\}] 
= - {1 \over 2} {r^2 [ \{\Oom_i\} ] \over u[\{\Oom_i\} ]},
\end{equation}
depending only on the positions $\{ \Oom_i \}$ of the zeros of $P ( \Oom
)$ where
\begin{eqnarray}
r[\{\Oom_i\}] & = & ( r - r_c ) \int d \Oom | P ( \Oom ) |^2 ,
\\
u[\{\Oom_i\}] & = & u \int d \Omega | P ( \Omega ) |^4 
- {4 \pi \over \kappa \Aa} \sum_{l,m} {1 \over
l (l^2 -1 ) ( l +2 )} | \Phi_{lm} |^2 . 
\nonumber
\end{eqnarray}
\par
The next step is to minimize $f[\{\Oom_i\}]$ over $\Oom_i$ 
Though in principle straightforward, this
task becomes quite complicated as $n$ grows large.  
In order to evaluate $f[\{\Oom_i\}]$, it is necessary to integrate $|P(\Oom
)|^2$ and $|P( \Oom )|^4$ over $\Oom$ and to calculate $\rho_{lm}$ for
arbitrary $\{\Oom_i \}$.  The function $|P ( \Oom )|^2$ can be expressed as
$\prod_i^{2n}(1 - \cos \gamma_i )$ where $\cos\gamma_i = \cos\theta
\cos\theta_i + \sin \theta \sin \theta_i \cos ( \phi - \phi_i)$.  Thus
$|P ( \Oom ) |^2$ is a polynomial of order $3^{2n}$ in
$\cos\theta$ and
$\sin\theta$.  $| \Phi ( \Oom ) |$ is more complicated; it is a sum of
$(n/2)(n+1)+1$ polynomials
of order $3^{2n}$.  We have been able to evaluate
$f( \{\Oom_i \})$ analytically for $n=1$ and 
$n=2$ (where $| P ( \Oom )|^2$ has $81$ terms).  For these two cases, we find,
as expected that the minimum energy configurations are, respectively, those
with zeros of $|P( \Oom ) |$ at opposite poles and at the vertices of a
tetrahedron.  For $n>2$, a complete evaluation of $f( \{ \Oom_i \})$ for
arbitrary $\{ \Oom_i \}$ becomes a daunting task. We,
therefore, appealed to symmetry to treat $n=3,4$ and $6$.  For $n=3$ and
$n=6$, we expect the zeros of $|P ( \Oom )|$
to lie, respectively, at the
vertices of an octahedron and an icosahedron.  For $n=4$, following the
calculations of Palthy-Muhoray[8],
we expect the zeros to lie at the vertices of a
distorted cube obtained by rotating its top face about its four-fold axis by
$\pi /4$ and compressing opposite faces.  For this case, we minimized the
energy over a single parameter determining the separation between opposite
rotated faces.
\par
The shape function $\rho ( \Oom ) = \psi_0^2 \rhob^{(n)}( \Oom )$ associated
with $n$-atic order is proportional to $\psi_0^2 \sim r - r_c$ to the order of
our calculations.  
In general, the Legendre decomposition
of $\rhob^{(n)}$ will contain Legendre polynomials of order $2n$.
For $n=1,2$, and $3$, we find
\begin{eqnarray}
\rhob^{(1)}(\Omega) & = & \frac{1}{4!}\frac{8}{3}\sqrt{\frac{\pi}{5}} Y_{2}^{0}
\nonumber \\
\rhob^{(2)}(\Omega) & = & \frac{1}{5!}\frac{80}{27}\sqrt{\frac{\pi}{7}}
   [Y_{3}^{0} + \sqrt{\frac{2}{5}}(-Y_{3}^{3}+Y_{3}^{-3})] 
                  +  \frac{2!}{6!}\frac{16}{45}\sqrt{\pi}
   [Y_{4}^{0} - \sqrt{\frac{10}{7}}(-Y_{4}^{3}+Y_{4}^{-3})] \label{shapes}\\
\rhob^{(3)}(\Omega) & = & \frac{2!}{6!}\frac{12}{11}\sqrt{\pi}[Y_{4}^{0} + 
               \sqrt{\frac{5}{14}}(Y_{4}^{4}+Y_{4}^{-4})]  
                  + \frac{4!}{8!}\frac{80}{3003}\sqrt{13\pi}
               [Y_{6}^{0} - \sqrt{\frac{7}{2}}(Y_{6}^{4}+Y_{6}^{-4})]
	       \nonumber
\end{eqnarray}
Fig. 1 shows the shapes described by these functions and the shapes $n=4$
and $n=6$ (whose shape functions are too long to display in this letter).
The transformations from the initial spherical shape to the distorted shapes
occur continuously. Our calculations for the shape are valid to order
$r-r_c$.  As temperature is lowered, higher order terms in $r-r_c$ and
higher order spherical harmonics are needed to describe the equilibrium
shape.  In Ref. [5], a variational function for $\psi$ and spherical
harmonics up to order
$8$ were used to calculate the shape for $n=1$ for
temperature well below the transition.  The shapes described by the
functions in Eq. (\ref{shapes}) exhibit unacceptable negative curvature
regions at
low temperature, and higher order spherical harmonics
(with the appropriate symmetry) are needed
to provide a correct description.
\par
We have presented here an analysis of the mean-field transition to $n$-atic
order on a fixed-area surface of genus zero.  
Our analysis is very similar to that of
Abrikosov for the transition from a normal metal to a vortex lattice in a
type II superconductor at $H_{c2}$ 
and completely ignores the effects of fluctuations, which may lead to
qualitative changes in our results.  We are really dealing with two kinds of
order: $n$-atic order and the positional order of vortices.  In mean-field
theory, these two kinds of order develop simultaneously.  In
superconductors, fluctuations drive the normal-to-superconducting transition
in a field first order[14].  In the Abrikosov phase, fluctuations of the vortex
lattice destroy[15] superconductivity but not long-range periodic order.  In
two-dimensions, screening of vortices drive the Kosterlitz-Thouless
transition in an infinite superconductor in zero field to zero[16].  
Both the above effects may be important for $n$-atic order on a
sphere, and we are currently investigating them. 
\acknowledgments
This work was supported in part by grants DMR-9122645 and DMR-8819885. 
We are grateful to Phil Nelson for a careful reading of the manuscript.
TCL and FCM are also grateful to Exxon Research and Engineering where a
portion of this work was carried out.

\figure{Mean-field shapes of deformable surfaces of genus zero 
with $n$-atic order.  Above the mean-field transition temperature, the
equilibrium shape is spherical for all $n$.  Below, the transition, the
equilibrium shape depends on $n$ and in general has a polyhedral form with
$2n$ vertices that coincide with the positions 
of strength $1/n$ vortices.}
\end{document}